\documentclass[twocolumn,aps,pra,groupedaddress,floatfix,nobalancelastpage]{revtex4-1}

\usepackage{graphicx}  
\usepackage{dcolumn}   
\usepackage{bm}        
\usepackage{amssymb}   
\usepackage{mathtools}
\usepackage[breaklinks=true]{hyperref}
\usepackage{amsthm}
\usepackage{ragged2e}
\usepackage{verbatim}
\usepackage{slashbox}
\usepackage{color,xcolor,colortbl}
\usepackage{soul}
\usepackage{multirow}
\usepackage{boxedminipage}
\usepackage{changepage}
\usepackage{bbm}
\usepackage{float}
\usepackage{array}
\usepackage{natbib}

\definecolor{shade}{gray}{0.90}

\newcommand{\bra}[1]{\langle #1|}
\newcommand{\ket}[1]{|#1\rangle}

\newcommand{\Tr}{\text{Tr}}

\newcommand{\sr}[2]{\rule[#1]{-2pt}{#2}}

\newcommand{\indices}[2]{\scalebox{0.85}{$\substack{#1\\#2}$}}

\begin{document}

\widetext

\title{Numerical evidence for bound secrecy from two-way postprocessing in quantum key distribution}
\author{Sumeet Khatri} \affiliation{Institute for Quantum Computing and the Department of Physics and Astronomy, University of Waterloo, Waterloo, ON, N2L 3G1}
\author{Norbert L\"{u}tkenhaus} \affiliation{Institute for Quantum Computing and the Department of Physics and Astronomy, University of Waterloo, Waterloo, ON, N2L 3G1}       

\date{\today}

\begin{abstract}
	
	Bound secret information is classical information that contains secrecy but from which secrecy cannot be extracted. The existence of bound secrecy has been conjectured but is currently unproven, and in this work we provide analytical and numerical evidence for its existence. Specifically, we consider two-way postprocessing protocols in prepare-and-measure quantum key distribution based on the well-known six-state signal states. In terms of the quantum bit-error rate $Q$ of the classical data, such protocols currently exist for $Q<\frac{5-\sqrt{5}}{10}\approx 27.6\%$. On the other hand, for $Q\geq\frac{1}{3}$ no such protocol can exist as the observed data is compatible with an intercept-resend attack. This leaves the interesting question of whether successful protocols exist in the interval $\frac{5-\sqrt{5}}{10}\leq Q<\frac{1}{3}$.
	
	Previous work has shown that a necessary condition for the existence of two-way postprocessing protocols for distilling secret key is breaking the symmetric extendability of the underlying quantum state shared by Alice and Bob. Using this result, it has been proven that symmetric extendability can be broken up to the $27.6\%$ lower bound using the advantage distillation protocol. In this work, we first show that to break symmetric extendability it is sufficient to consider a generalized form of advantage distillation consisting of one round of postselection by Bob on a block of his data. We then provide evidence that such generalized protocols cannot break symmetric extendability beyond $27.6\%$. We thus have evidence to believe that $27.6\%$ is an upper bound on two-way postprocessing and that the interval $\frac{5-\sqrt{5}}{10}\leq Q<\frac{1}{3}$ is a domain of bound secrecy.
	 
\end{abstract}

\maketitle

\section{Introduction}	
	
	When considering prepare-and-measure (PM)-based quantum key distribution (QKD) protocols, to what extent can an eavesdropper tamper with the signals being sent by the sender (Alice) to the receiver (Bob) before classical postprocessing protocols on the resulting measurement data cannot distill a secret key? More generally, which bipartite probability distributions between Alice and Bob contain secret bits that can be distilled into a secret key by some classical protocol?
	
	Suppose Alice and Bob's data are given by measurement of many copies of a given quantum state $\rho^{AB}$. If $\rho^{AB}$ is separable, then it is known that an intercept-resend attack exists and therefore no classical postprocessing protocol can distill a secret key \cite{CurtyEntanglement}. If $\rho^{AB}$ is \textit{symmetrically extendable to a copy of $B$}, then there exists a tripartite extension $\rho^{ABB'}$ of $\rho^{AB}$ such that $\rho^{AB'}=\rho^{AB}$ \footnote{See \cite{MyhrThesis} for an introduction to symmetrically extendable states. See \cite{ChenSymext,RanadeSymExt} for the only currently known necessary and sufficient criteria for symmetric extendability that hold for two classes of states. From now on, by symmetrically extendable we will always mean symmetrically extendable to a copy of Bob's system.}. In this situation, it is known that no one-way protocol involving communication from Alice to Bob can be used to distill a secret key because the system $B'$ is effectively a copy of $B$ and could belong to an eavesdropper (Eve), meaning that from Alice's point of view Bob and Eve are symmetric \cite{MoroderOneWay}.
	
\begin{figure}[t]
		\centering
		\includegraphics[scale=0.8]{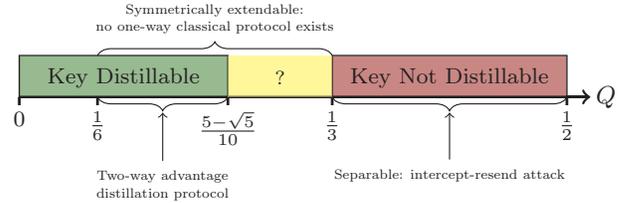}
		\caption{Key distillability as a function of the QBER $Q$ characterizing the quantum state $\rho_Q^{AB}$ from which Alice and Bob obtain their classical data by measurement in the standard basis. We are interested in whether a secret key can be distilled in the \textit{gap}, that is, the yellow region.}\label{fig-1}
	\end{figure}

	In this paper, we assume that Alice and Bob share many copies of the state $\rho_Q^{AB}=(1-2Q)\ket{\Phi^+}\bra{\Phi^+}+\frac{Q}{2}\mathbbm{1}_{AB}$, where $Q\in\left[0,\frac{1}{2}\right]$ is the quantum bit-error rate (QBER) and $\ket{\Phi^+}=\frac{1}{\sqrt{2}}(\ket{0,0}+\ket{1,1})$. They obtain their classical data by local measurement of each state in the standard basis $\{\ket{0},\ket{1}\}$. This situation arises in PM-based six-state QKD protocols \cite{BrussSix-State}. Since such protocols are tomographically complete, under a collective attack and in the infinite key limit, Alice and Bob can verify that they indeed hold many copies of the state $\rho_Q^{AB}$ \cite{MyhrThesis}. As indicated in Fig. \ref{fig-1}, it is known that $\rho_Q^{AB}$ is separable for $Q\geq\frac{1}{3}$ and symmetrically extendable for $Q\geq\frac{1}{6}$ \cite{MoroderOneWay}. By the arguments above, distilling secret key beyond $\frac{1}{6}$ would require a \textit{two-way} classical postprocessing protocol.
	
	Gottesman and Lo \cite{GottesmanLo} were the first to examine two-way classical postprocessing protocols for PM-based QKD protocols. For the scenario we are considering here, they devised a two-way protocol that could distill secret key up to $Q=26.4\%$. Chau \cite{Chau} modified the Gottesman-Lo protocol and showed that a secret key could be distilled up to $Q=\frac{5-\sqrt{5}}{10}\approx 27.6\%$, as shown in Fig. \ref{fig-1}. The protocols used by Gottesman and Lo and Chau were entanglement distillation protocols adapted to hold in a classical setting based on the Shor-Preskill proof of security of the BB84 protocol \cite{BB84,ShorPreskill_BB84}. The corresponding purely classical protocol is advantage distillation, which is a two-way classical protocol devised by Maurer \cite{MaurerAD} in which Alice and Bob postselect on the repetition codes $\mathcal{R}_n=\{00\dotsb 00,11\dotsb 11\}$ for some block length $n$. Ac\'{\i}n \cite{Acin1} and Bae and Ac\'{\i}n \cite{BaeAcin} showed explicitly that advantage distillation followed by error correction and privacy amplification could distill secret key up to the Chau threshold and that beyond this threshold there exists an eavesdropping attack that causes advantage distillation to fail. To try to distill secret key beyond the Chau threshold, i.e., in the gap, Bae and Ac\'{\i}n allowed Alice and Bob to perform noisy preprocessing before advantage distillation, and they even allowed Bob to perform coherent operations. Neither of these modifications could distill secret key beyond the Chau threshold.
	
	Myhr et al. \cite{MyhrPaper} then provided a different perspective on two-way protocols by shifting the goal from distilling a secret key to \textit{breaking symmetric extendability}. Specifically, they argued that if Alice and Bob's initial data corresponds to a symmetrically extendable state, then any successful two-way protocol must first transform this state to one that is not symmetrically extendable. This is due to the fact that any two-way protocol necessarily ends with a final round of one-way communication, which cannot be successful in distilling a secret key unless Alice and Bob's correlations are not symmetrically extendable. They showed that all previous work pertaining to two-way protocols beyond $Q=\frac{1}{6}$ could be viewed in this way. In particular, they showed that advantage distillation breaks symmetric extendability up to the Chau threshold and that the existence of an eavesdropping attack beyond this threshold corresponds to the existence of a symmetric extension of the effective quantum state after advantage distillation. Bae and Ac\'{\i}n's unsuccessful attempts to go beyond the Chau threshold can be understood similarly since local quantum operations, in particular noisy preprocessing and coherent operations by Bob, preserve symmetric extendability. To break symmetric extendability, Myhr et al. proved that it is sufficient to consider one announcement by Bob to Alice on a block of his data that can be described by one Kraus operator on the quantum states. To this end, Myhr \cite{MyhrThesis} considered a generalized form of advantage distillation in which Alice and Bob postselect on some pre-chosen linear error correction code. He provided analytical and numerical evidence to suggest that such generalized protocols could not break symmetric extendability beyond the Chau threshold. Since only postselection on linear codes was considered, his results left open the possibility that postselection on \textit{nonlinear} codes might be able to break symmetric extendability beyond the Chau threshold.
	
	We now consider postselection on nonlinear codes using the single-Kraus-operator formulation from Myhr et al. Specifically, in Sec. \ref{sec-formulation} we provide an explicit form for the Kraus operator, which allows us to obtain the effective quantum state after postselection by Bob on \textit{arbitrary} error correction codes, i.e., both linear and nonlinear codes. We describe the structure of the effective states and show that the search for codes breaking symmetric extendability beyond the Chau threshold can be reduced to the search over inequivalent codes. In Sec. \ref{sec-thresholds}, we show analytically that repetition codes achieve the Chau threshold and numerically determine updated thresholds for inequivalent codes of small block lengths and number of code words. From these results, we observe that repetition codes are optimal for each block length. In Sec. \ref{sec-Alice_postselection}, we consider postselection by Alice and Bob on the same code. From a random search on over 540,000 codes, we find that none are able to break symmetric extendability beyond the Chau threshold. We also introduce a procedure to construct a symmetric extension that works for 99\% of the tested codes.  
	
	Our results lead us to the conjecture that repetition codes are optimal for each block length, meaning that there does \textit{not} exist a code that can break symmetric extendability beyond the Chau threshold and therefore that a secret key cannot be distilled in the gap. Since in the gap $\rho_Q^{AB}$ is entangled, the corresponding classical data contains secret bits \cite{Acin3}. Therefore, if our conjecture is true, the data corresponding to the gap would contain \textit{bound secrecy}, the classical analogue of bound entanglement in which classical data contain secret bits that cannot be extracted into a secret key by any protocol. The gap itself would then be an example of the separation between secrecy formation and secrecy extraction, the classical analogue of the separation between entanglement of formation and distillable entanglement. The existence of such a separation, as well as the existence of bound secrecy, has been conjectured with much evidence for its existence \cite{Gisin2000,Gisin2001,Gisin2002,Entanglement_classical_analog,RenWol03,Acin3}, but a proof has still not been found. Much of the prior evidence has focused on classical data arising from measurement of bound entangled states. Our results suggest that bound secrecy can be obtained even from measurement of quantum states with distillable entanglement.

\section{Formulation of the Problem}\label{sec-formulation}

	Even though PM-based protocols never actually involve bipartite entangled states, all such protocols can be modelled mathematically in terms of Alice preparing some pure bipartite entangled state $\ket{\psi}^{AB}$, measuring one half of it, and sending the other half to Bob, who then measures \cite{LutkenhausChapter}. The resulting classical measurement data then corresponds to the quantum state $\rho^{AB}\coloneqq(\mathbbm{1}_A\otimes \Phi)(\ket{\psi}\bra{\psi}^{AB})$ through the distribution $p_{AB}(i,j)=\Tr[(A_i\otimes B_j)\rho^{AB}]$, where $\{A_i\}_i$ and $\{B_j\}_j$ are Alice and Bob's measurement positive operator-valued measures (POVMs) and $\Phi$ is the channel through which Alice sends her signals to Bob and is assumed to be under Eve's control.
	
	We are considering in this paper PM-based six-state QKD protocols in which Alice and Bob's classical data arise from individual measurement in the standard basis of several copies of the state $\rho_Q^{AB}=(1-2Q)\ket{\Phi^+}\bra{\Phi^+}+\frac{Q}{2}\mathbbm{1}_{AB}$. For $Q\geq\frac{1}{6}$, which is when this state is symmetrically extendable, no one-way classical postprocessing protocol can distill a secret key, which means we must consider two-way protocols. We know from Myhr et al. \cite{MyhrPaper} that any successful two-way protocol must transform the symmetrically extendable state $\rho_Q^{AB}$ to an updated state that is not symmetrically extendable. They also proved that if we care only about breaking symmetric extendability (and not, say, about the value of the success probability and/or the secret-key rate), then it is sufficient to consider the symmetric extendability of the updated states corresponding to a single announcement by Bob on a block of his data of length $n$ that can be described by one Kraus operator. If no such announcement can break symmetric extendability, then no general announcement will be able to, hence no two-way protocol will be able to distill a secret key. Since Bob's data are purely classical, his announcement can only be based on some partitioning of the $n$-bit strings $\{0,1\}^n$. Suppose the set $\mathcal{C}=\{C_k\}_{k=0}^{m-1}$ of $m$ distinct $n$-bit strings is one of the partitions. If Bob announces that his block of data is in $\mathcal{C}$, then the updated quantum state consistent with Alice and Bob's updated correlations, and the one from which key distillability after Bob's announcement can be determined, is 
	\begin{equation}\label{eq-filtered_state}
		\rho_{Q,\mathcal{C}}^{A^n\tilde{B}}=(\mathbbm{1}_{A^n}\otimes K_{\mathcal{C}})(\rho_Q^{AB})^{\otimes n}(\mathbbm{1}_{A^n}\otimes K_{\mathcal{C}})^\dagger,
	\end{equation}
	where 
	\begin{equation}\label{eq-filter}
		K_{\mathcal{C}}=\sum_{k=0}^{m-1}\ket{k}\bra{C_k}
	\end{equation}
	is the single Kraus operator corresponding to Bob's announcement. This is due to the fact that the conditional probability distribution of Alice and Bob's data given that Bob's data are in $\mathcal{C}$ is the same as the probability distribution of Alice measuring the updated state \eqref{eq-filtered_state} in the $n$-qubit standard basis and Bob measuring the updated state in the basis $\{\ket{k}\}_{k=0}^{m-1}$ corresponding to the elements of $\mathcal{C}$. Note that we get the same quantum state \eqref{eq-filtered_state} if Bob merely performs \textit{postselection} on the set $\mathcal{C}$, that is, if Bob keeps his block of data if it is in $\mathcal{C}$ and discards it otherwise, publicly announcing in each case what he does. In this way, we obtain a generalized form of the advantage distillation protocol.
	
	Now, to determine the existence of two-way protocols, it is sufficient to consider the symmetric extendability of states of the form \eqref{eq-filtered_state} \footnote{Normalization of the state is unimportant for symmetric extendability since if $\rho^{AB}$ is symmetrically extendable then so is $\alpha\rho^{AB}$ for any $\alpha>0$.}. The reason for this is that any general announcement scheme will be based on accepting the block of data if it is in some partition(s) of the $n$-bit strings. If announcing on each partition alone (for which the updated state is of the form \eqref{eq-filtered_state}) cannot break symmetric extendability, neither can announcing on multiple partitions. We are thus interested in the symmetric extendability of the states \eqref{eq-filtered_state}. Specifically, we are interested in the \textit{updated threshold} $Q_{\mathcal{C}}^*$, which we define as the value of the QBER beyond which the state is symmetrically extendable. (Note that without the Kraus operator the threshold is simply $\frac{1}{6}$ for all $n$.) The problem is then to find a set $\mathcal{C}$ with a threshold exceeding the Chau threshold, i.e., a set that breaks symmetric extendability in the gap. 
	
	Being a subset of the $n$-bit strings, the set $\mathcal{C}$ can be thought of as a (classical) error correction code of block length $n$ with the elements of the set being the code words. We will therefore throughout the rest of the paper call the set on which Bob postselects a code and the elements of the set the code words. We will call $\mathcal{C}$ an \textit{$(n,m)$ code} if it has block length $n$ and contains $m$ code words. $\mathcal{C}$ is called \textit{linear} if it is closed under the bitwise XOR operation $\oplus$, and \textit{nonlinear} otherwise. Note that linear codes necessarily have $2^k$ code words for some $k$.

\subsection{Structure of the Updated States}\label{subsec-structure_filtered_states}
	
	The updated state \eqref{eq-filtered_state} can be written as
	\begin{equation*}
		\rho_{Q,\mathcal{C}}^{A^n\tilde{B}}=\sum_{\alpha,\alpha'\in\{0,1\}^n}\sum_{k,k'=0}^{m-1} \left(\rho_{Q,\mathcal{C}}^{A^n\tilde{B}}\right)_{\indices{\alpha,k}{\alpha',k'}}\ket{\alpha,k}\bra{\alpha',k'},
	\end{equation*}
	where \cite{KhatriThesis}
	\begin{equation}\label{eq-filtered_state_matrix_elements}
		\begin{aligned}
		\left(\rho_{Q,\mathcal{C}}^{A^n\tilde{B}}\right)_{\indices{\alpha,k}{\alpha',k'}}&=\left(\frac{1-2Q}{2}\right)^{|\alpha\oplus\alpha'|}\left(\frac{Q}{2}\right)^{|\alpha\oplus C_{k}|}\\
		&\quad\times\left(\frac{1-Q}{2}\right)^{n-|\alpha\oplus\alpha'|-|\alpha\oplus C_{k}|}\\
		&\quad\times\delta_{\alpha\oplus C_{k},\alpha'\oplus C_{k'}}\delta_{(\alpha\oplus C_{k})\odot(\alpha\oplus\alpha'),\underline{0}^n}.
		\end{aligned}
	\end{equation}
	Here, $\odot$ is the bit-wise AND operation, and $|\alpha|$ is the Hamming weight (number of ones) of the bit string $\alpha$. The representation of the updated state in this manner depends on the ordering of $\mathcal{C}$ and $\{0,1\}^n$. A different ordering of these sets changes the updated state by local unitaries, which does not affect the corresponding threshold $Q_{\mathcal{C}}^*$ since symmetrically extendable states are preserved under local unitaries. We therefore consider updated states corresponding to the same code to be equal if they differ only by local unitaries corresponding to a different ordering of the code words.
	
	The Kronecker delta $\delta_{\alpha\oplus C_k,\alpha'\oplus C_{k'}}$ in Eq. \eqref{eq-filtered_state_matrix_elements} indicates that the elements of the updated state are nonzero if and only if $\alpha\oplus C_k=\alpha'\oplus C_{k'}$. This means that the (ordered) basis $\{\ket{\alpha,k}:\alpha\in\{0,1\}^n,~0\leq k\leq m-1\}$ can be changed to the new (ordered) basis $\bigcup_{\beta\in\{0,1\}^n}\{\ket{\alpha,k}:\alpha\oplus C_k=\beta\}$ by a unitary $V$ that leaves the updated state in a \textit{block-diagonal} form,
	\begin{equation}\label{eq-filtered_state_blkdiag}
		V\rho_{Q,\mathcal{C}}^{A^n\tilde{B}}V^\dagger=\bigoplus_{\beta\in\{0,1\}^n} M_{Q,\mathcal{C}}^{(\beta)},
	\end{equation}
	where $M_{Q,\mathcal{C}}^{(\beta)}$ are the blocks, each $m\times m$, with elements $\left(M_{Q,\mathcal{C}}^{(\beta)}\right)_{\indices{k}{k'}}=\left(\rho_{Q,\mathcal{C}}^{A^n\tilde{B}}\right)_{\indices{\beta\oplus C_k,k}{\beta\oplus C_{k'},k'}}$ for all $0\leq k,k'\leq m-1$. 
	
	For a \textit{direct sum} of codes $\mathcal{C}_1=\{C_{1,k}\}_{k=0}^{m_1-1}$ of block length $n_1$ and $\mathcal{C}_2=\{C_{2,k}\}_{k=0}^{m_2-1}$ of block length $n_2$, the updated state has a tensor product structure in addition to the block-diagonal structure. The direct sum is the code $|\mathcal{C}_1|\mathcal{C}_2|=\{C_{1,k}C_{2,\ell}:0\leq k\leq m_1-1,~0\leq \ell\leq m_2-1\}$ with block length $n_1+n_2$ and $m_1m_2$ code words \cite{SloneMacWilliams}. In this case, it is straightforward to show \cite{KhatriThesis} that the updated state is equal to
	\begin{equation}\label{eq-filtered_state_direct_sum}
		\rho_{Q,|\mathcal{C}_1|\mathcal{C}_2|}^{A^{n_1+n_2}\tilde{B}}=\rho_{Q,\mathcal{C}_1}^{A^{n_1}\tilde{B}_1}\otimes\rho_{Q,\mathcal{C}_2}^{A^{n_2}\tilde{B}_2}.
	\end{equation}
	The corresponding threshold is then equal to
	\begin{equation}\label{eq-threshold_direct_sum}
		Q_{|\mathcal{C}_1|\mathcal{C}_2|}^*=\max\{Q_{\mathcal{C}_1}^*,Q_{\mathcal{C}_2}^*\}
	\end{equation}
	since the state is symmetrically extendable if and only if each state in the tensor product is symmetrically extendable. Analogous results for the updated state and threshold hold for a direct sum of more than two codes, which is defined analogously to the direct sum of two codes.

\subsection{Equivalent Codes}\label{subsec-equivalent_codes}
	
	Two codes $\mathcal{C}=\{C_k\}_{k=0}^{m-1}$ and $\mathcal{D}=\{D_k\}_{k=0}^{m-1}$, each with block length $n$, are called \textit{equivalent} if there exists a permutation $\pi$ on $n$ bits and $\alpha\in\{0,1\}^n$ such that $\pi(\mathcal{D})\oplus\alpha\coloneqq\{\pi(D_k)\oplus\alpha:0\leq k\leq m-1\}=\mathcal{C}$ \cite{SloneMacWilliams}. If $\mathcal{C}$ and $\mathcal{D}$ are equivalent, then \cite{KhatriThesis}
	\begin{equation}\label{eq-equivalence}
		\rho_{Q,\mathcal{C}}^{A^n\tilde{B}}=\rho_{Q,\mathcal{D}}^{A^n\tilde{B}}\quad\forall~0\leq Q\leq\frac{1}{2},
	\end{equation}
	which means that $Q_{\mathcal{C}}^*=Q_{\mathcal{D}}^*$. Therefore, for any given block length $n$ and number of code words $m$, to find a code exceeding the Chau threshold we need only search the inequivalent codes.
	
	We can remove from the set of inequivalent codes those that contain constant columns when written as a matrix with each code word forming a row. This is due to the fact that any such code is equivalent to a code of the form $\mathcal{C}'=\{\beta C_k\}_{k=0}^{m-1}=|\{\beta\}|\mathcal{C}|$ for some $\beta\in\{0,1\}^x$, and by Eq. \eqref{eq-filtered_state_direct_sum} it holds that
	\begin{equation}
		\rho_{Q,\mathcal{C}'}^{A^{n+x}\tilde{B}}=\rho_{Q,\{\beta\}}^{A^x\tilde{B}_1}\otimes\rho_{Q,\mathcal{C}}^{A^n\tilde{B}_2}.
	\end{equation}
	Since $\rho_{Q,\{\beta\}}^{A^x\tilde{B}_1}$ is symmetrically extendable for all $Q\in\left[0,\frac{1}{2}\right]$ ($\tilde{B}_1$ is one-dimensional), it holds that $Q_{\mathcal{C}'}^*=Q_{\mathcal{C}}^*$. In other words, removing the constant columns gives a code without constant columns (and with smaller block length) that has the same threshold as the original code.

	For any given $(n,m)$ class, we can thus take all codes without constant columns and run through all pairs of permutations and bit strings to determine the number of inequivalent codes. For small $(n,m)$ classes, we obtain the numbers in Table \ref{table-inequivalent_sets}. Notably, with two code words, the repetition code $\mathcal{R}_n=\{00\dotsb 00,11\dotsb 11\}$ is the only inequivalent code without constant columns for all $n\geq 2$. Determining the number of inequivalent codes for higher $(n,m)$ classes becomes very rapidly lengthy since the number of permutations is $n!$ and the number of bit strings is $2^n$, resulting in $n!\times 2^n$ pairs of permutations and bit strings over which to search.
	
	\begin{table}[t]
		\centering
		\begin{tabular}{|c||c|c|c|c|c|c|c|c|c|c|c|c|c|c|}
			\hline\backslashbox{$n$}{$m$} & 2 & 3 & 4 & 5 & 6 & 7 & 8 & 9 & 10 & 11 & 12 & 13 & 14 & 15 \\ \hline\hline
				2 & 1 & 1 & \cellcolor{shade} & \cellcolor{shade} & \cellcolor{shade} & \cellcolor{shade} & \cellcolor{shade} & \cellcolor{shade} & \cellcolor{shade} & \cellcolor{shade} & \cellcolor{shade} & \cellcolor{shade} & \cellcolor{shade} & \cellcolor{shade}  \\ \hline
				3 & 1 & 2 & 5 & 3 & 3 & 1 & \cellcolor{shade} & \cellcolor{shade} & \cellcolor{shade} & \cellcolor{shade} & \cellcolor{shade} & \cellcolor{shade} & \cellcolor{shade} & \cellcolor{shade}  \\ \hline
				4 & 1 & 3 & 13 & 24 & 47 & 55 & 73 & 56 & 50 & 27 & 19 & 6 & 4 & 1  \\ \hline
				5 & 1 & 4 & 28 & 104 & 422 & $\cdot$ & $\cdot$ & $\cdot$ & $\cdot$ & $\cdot$ & $\cdot$ & $\cdot$ & $\cdot$ & $\cdot$ \\ \hline
				6 & 1 & 6 & 56 & $\cdot$ & $\cdot$ & $\cdot$ & $\cdot$ & $\cdot$ & $\cdot$ & $\cdot$ & $\cdot$ & $\cdot$ & $\cdot$ & $\cdot$ \\ \hline
				7 & 1 & 7 & $\cdot$ & $\cdot$ & $\cdot$ & $\cdot$ & $\cdot$ & $\cdot$ & $\cdot$ & $\cdot$ & $\cdot$ & $\cdot$ & $\cdot$ & $\cdot$ \\ \hline
				8 & 1 & 9 & $\cdot$ & $\cdot$ & $\cdot$ & $\cdot$ & $\cdot$ & $\cdot$ & $\cdot$ & $\cdot$ & $\cdot$ & $\cdot$ & $\cdot$ & $\cdot$ \\ \hline
		\end{tabular}
		\caption{Number of inequivalent codes without constant columns for some small $(n,m)$ classes. For each $n$, the largest $m$ such that the code is non-trivial is $2^n-1$. Cells beyond this value have thus been shaded. Numbers for the cells with a dot have not been calculated.}\label{table-inequivalent_sets}
	\end{table}

\section{Updated Thresholds}\label{sec-thresholds}
	
	Our goal is to find a code such that the corresponding updated state \eqref{eq-filtered_state} has a threshold in the gap of Fig. \ref{fig-1}. From the previous section, we know that it is sufficient to search over inequivalent codes without constant columns. In this section, we determine the threshold of the codes in many of the $(n,m)$ classes for which the number of inequivalent codes without constant columns has been determined, as indicated in Table \ref{table-inequivalent_sets}, where $n$ is the block length and $m$ the number of code words. First, in Sec. \ref{subsec-repetition_codes}, we provide the analytical result that repetition codes, the only inequivalent code without constant columns containing two code words, reproduce the Chau threshold and the known result from \cite{MyhrPaper} that advantage distillation cannot break symmetric extendability in the gap. In fact, our result builds on that result from \cite{MyhrPaper} since in that work they considered postselection by Alice and Bob on repetition codes while we consider postselection only by Bob. (We consider the less general case of Alice postselecting on the same code as Bob in Sec. \ref{sec-Alice_postselection}.) We then numerically determine the thresholds for other $(n,m)$ classes in Sec. \ref{subsec-numerics_thresholds} and describe trends in the thresholds that lead us to believe that repetition codes are optimal for each block length. In Sec. \ref{subsec-best_codes}, we examine the best codes among the tested $(n,m)$ classes and in Sec. \ref{subsec-future_work} we go through some areas for future work.
	
\subsection{Repetition Codes}\label{subsec-repetition_codes}

	For the repetition codes $\mathcal{R}_n=\{00\dotsb 00,11\dotsb 11\}$, it holds that for each $n\geq 1$ the threshold $Q_{\mathcal{R}_n}^*$ is the solution to
	\begin{equation}\label{eq-repetition_symext_equation}
		4Q^{2n}-4Q^n(1-Q)^n+(1-2Q)^{2n}=0.
	\end{equation}
	We provide a sketch of the proof of this fact in Appendix \ref{appendix-repetition_code_threshold}, while the full proof can be found in \cite{KhatriThesis}. 
	
\begin{figure}[t]
		\centering
		\includegraphics[scale=0.40]{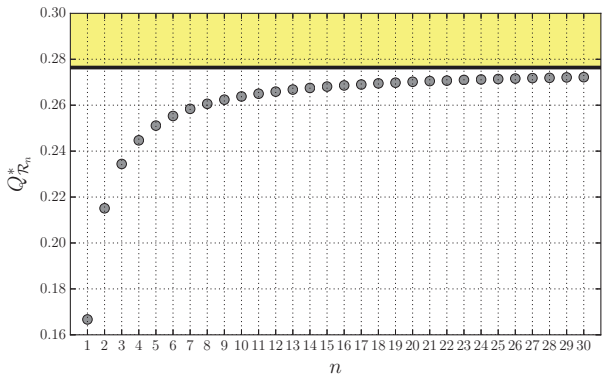}
		\caption{Repetition code thresholds $Q_{\mathcal{R}_n}^*$ up to $n=30$.}\label{fig-AD_thresholds}
	\end{figure}

	The thresholds $Q_{\mathcal{R}_n}^*$ are plotted in Fig. \ref{fig-AD_thresholds}. They increase monotonically with $n$ and appear to converge to the Chau threshold $\frac{5-\sqrt{5}}{10}\approx 27.6\%$ in the limit $n\rightarrow\infty$. Indeed, it can be shown \cite{KhatriThesis} that as $n$ increases the thresholds approach
	\begin{equation}
		\scalebox{0.85}{$\tilde{Q}_{\mathcal{R}_n}^*\coloneqq \frac{\left(4\left(\frac{1}{4}\right)^{\frac{1}{n}}+1\right)-\sqrt{\left(4\left(\frac{1}{4}\right)^{\frac{1}{n}}+1\right)^2-4\left(\frac{1}{4}\right)^{\frac{1}{n}}\left(4\left(\frac{1}{4}\right)^{\frac{1}{n}}+1\right)}}{2\left(4\left(\frac{1}{4}\right)^{\frac{1}{n}}+1\right)}.$}
	\end{equation}
	Since $\lim_{n\rightarrow\infty}\tilde{Q}_{\mathcal{R}_n}^*=\frac{5-\sqrt{5}}{10}$, which is precisely the Chau threshold, it holds that the actual threshold approaches the Chau threshold as well. This reproduces the result from \cite{MyhrPaper} that advantage distillation cannot break symmetric extendability beyond the Chau threshold, although, as mentioned earlier, we have considered postselection only by Bob while in \cite{MyhrPaper} postselection by Alice as well was considered.
	
\subsection{Numerical Determination of Thresholds for Arbitrary Codes}\label{subsec-numerics_thresholds}
	
	There are currently no known symmetric extendability results that allow us to determine the thresholds for codes other than repetition codes. Fortunately, we can still numerically determine the thresholds using the following semidefinite program (SDP) that can be used to determine the symmetric extendability of an arbitrary positive semidefinite operator $P^{AB}$:
	\begin{equation}\label{eq-symext_SDP}
	\begin{array}{l l}
		\text{min.} & t\\
		\text{subject to} & R^{ABB'}+t\mathbbm{1}^{ABB'}\geq 0,\\
			& \Tr_{B}[R^{ABB'}]=P^{AB},\\
			& \Tr_{B'}[R^{ABB'}]=P^{AB}.
	\end{array}
	\end{equation}
	When the minimum value $t_{\text{min}}(P^{AB})$ of the objective function $t$ is positive $P^{AB}$ is not symmetrically extendable, and when $t_{\text{min}}(P^{AB})$ is nonpositive $P^{AB}$ is symmetrically extendable, with the operator $R^{ABB'}$ achieving the minimum being a symmetric extension.
	
\begin{figure}[t]
		\centering
		\includegraphics[scale=0.7]{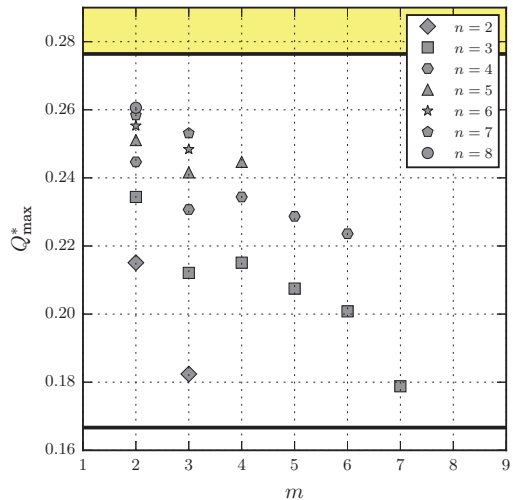}
		\caption{The highest thresholds for some of the $(n,m)$ classes from Table \ref{table-inequivalent_sets}.}\label{fig-plot_thresholds}
	\end{figure}

	For the updated states \eqref{eq-filtered_state}, each code $\mathcal{C}$ corresponds to the function $T_\mathcal{C}:\left[0,\frac{1}{2}\right]\rightarrow \mathbbm{R}$ defined by
	\begin{equation}\label{eq-threshold_function}
		T_{\mathcal{C}}(Q)=t_{\text{min}}(\rho_{Q,\mathcal{C}}^{A^n\tilde{B}}).
	\end{equation}
	The zero of $T_{\mathcal{C}}$ is then the threshold $Q_{\mathcal{C}}^*$. To obtain the threshold, we selected 200 points (evenly spaced) in the QBER interval $\mathcal{I}=[0.16,0.33]$ and determined the zero of the curve of best fit to the points $\{T_{\mathcal{C}}(Q):Q\in\mathcal{I}\}$ corresponding to an estimate of the function $T_{\mathcal{C}}$. The points $T_{\mathcal{C}}(Q)$ were obtained by solving the SDP \eqref{eq-symext_SDP} in {\scshape{Matlab}} using {\scshape{yalmip}} \cite{yalmip} with the solver {\scshape{SCS}} \cite{SCS_solver} to a precision of $10^{-10}$. The best-fitting curve was obtained in {\scshape{Matlab}} using a cubic spline model. This procedure leads to thresholds that are accurate to within $8.5\times 10^{-4}$, which is the spacing between the points in the interval $\mathcal{I}$. For codes containing more than two code words ($m>2$), we applied this procedure to many of the $(n,m)$ classes in Table \ref{table-inequivalent_sets} for which the number of inequivalent codes without constant columns is known. Specifically, for $m=3$ we determined the thresholds up to $n=7$, for $m=4$ up to $n=5$, for $m=5$ and $m=6$ up to $n=4$, and for $m=7$ up to $n=3$. The highest threshold for each $(n,m)$ class tested is plotted in Fig. \ref{fig-plot_thresholds}. Determining the threshold for higher $(n,m)$ classes becomes increasingly time consuming and resource intensive since each code requires 200 SDPs and each SDP takes longer to complete, and requires more computer memory, as $n$ and $m$ increase and with it the number of optimization variables. As well, since the dimension of Alice's space is $2^n$, memory limits can be quickly reached in just storing the updated state. 
	
	From the plot in Fig. \ref{fig-plot_thresholds}, we observe that for each block length $n$, the $(n,2)$ class, which contains only the repetition code $\mathcal{R}_n$, has the highest threshold. For fixed $n$, as the number of code words $m$ increases beyond 2, the threshold tends to decrease. This suggests that repetition codes are optimal for any given block length and therefore that codes with a high number of code words are unlikely to be helpful in obtaining a threshold exceeding the Chau threshold. In particular, we have a clear sign that nonlinear codes are not necessarily better than linear codes and that they are actually worse than linear codes in some cases since the $m=3,5,6,7$ codes are necessarily nonlinear and their highest thresholds are less than the highest thresholds of the $m=2$ and $m=4$ codes, which are given by linear codes. (We see that the best $m=4$ codes are linear in the next section.) As well, for fixed $m$, the threshold increases with increasing $n$, with a clear indication, particularly for small $m$, that the thresholds are converging to the Chau threshold.

\subsection{Codes with the Highest Threshold}\label{subsec-best_codes}

	Having seen that the repetition code appears optimal for each block length $n$, we now examine the best codes, i.e., the codes with the highest threshold, for each number of code words $m$. Since for all block lengths the repetition code is the only inequivalent code without constant columns containing two code words, it is the best code with two code words. In the following, we write codes as a matrix in which each row of the matrix is a code word.
	
	With three code words, the best codes are
	\begin{equation}\label{eq-best_three_codewords}
		\begin{aligned}
		\left[\begin{array}{c}00\\01\\11\end{array}\right],&~\left[\begin{array}{c} 000\\011\\111\end{array}\right],~\left[\begin{array}{c}0000\\0111\\1111\end{array}\right],~\left[\begin{array}{c}00000\\01111\\11111\end{array}\right],\\
		&\left[\begin{array}{c}000000\\011111\\111111\end{array}\right],~\left[\begin{array}{c}0000000\\0111111\\1111111\end{array}\right].
		\end{aligned}
	\end{equation}
	There is a clear trend among these codes, so we conjecture that for all $n\geq 2$ the best three-code-word code is
	\begin{equation*}
		\left[\begin{array}{c} 00\dotsb 00\\ 01\dotsb 11\\ 11\dotsb 11\end{array}\right].
	\end{equation*}
	
	With four code words, the best codes are
	\begin{equation}\label{eq-best_four_codewords}
		\left[\begin{array}{c}000\\011\\100\\111\end{array}\right],~\left[\begin{array}{c} 0000\\0111\\1000\\1111\end{array}\right],~\left[\begin{array}{c} 00000\\01111\\10000\\11111\end{array}\right].
	\end{equation}
	Each of these codes is linear and equal to $|\mathcal{R}_1|\mathcal{R}_{n-1}|$, so that the threshold is equal to $Q_{\mathcal{R}_{n-1}}^*$. We conjecture therefore that for all $n\geq 3$ the best four-code-word code is $|\mathcal{R}_1|\mathcal{R}_{n-1}|$.
	
	With five code words, the best codes are
	\begin{equation}\label{eq-best_five_codewords}
		\left[\begin{array}{c}000\\001\\011\\100\\111\end{array}\right],~\left[\begin{array}{c}0000\\0011\\0111\\1000\\1111\end{array}\right].
	\end{equation}
	Both of these codes are of the form
	\begin{equation}\label{eq-best_five_codewords_conjecture}
		\left[\begin{tabular}{c|c}
			0 & $00\dotsb 00$\\
			0 & $01\dotsb 11$\\
			0 & $11\dotsb 11$\\ \hline
			1 & $00\dotsb 00$\\
			1 & $11\dotsb 11$\\ 
		\end{tabular}\right].
	\end{equation}
	That is, the two codes are composed of the code words of the conjectured best $(n-1,3)$ code and the code words of the repetition code $\mathcal{R}_{n-1}$. As well, the thresholds of the two codes are between the thresholds of the two smaller codes comprising them.
	
	With six code words, the best codes are
	\begin{equation}\label{eq-best_six_codewords}
		\left[\begin{array}{c}000\\001\\010\\011\\100\\111\end{array}\right],~\left[\begin{array}{c}0000\\0011\\0101\\0111\\1000\\1111\end{array}\right].
	\end{equation}
	Like the best five-code-word codes in Eq. \eqref{eq-best_five_codewords}, these codes are of the form
	\begin{equation}\label{eq-best_six_codewords_conjecture}
		\left[\begin{tabular}{c|c}
			0 & \multirow{4}{1cm}{\centering$\mathcal{C}$}\\
			0 & \\
			0 & \\
			0 & \\ \hline
			1 & \multirow{2}{1cm}{\centering$\mathcal{R}_{n-1}$}\\
			1 & \\ 
		\end{tabular}\right]
	\end{equation}
	for an $(n-1,4)$ code $\mathcal{C}$. In Eq. \eqref{eq-best_five_codewords}, $\mathcal{C}$ is the best code in the $(n-1,3)$ class, while in the best $(4,6)$ code $\mathcal{C}$ is \textit{not} the best $(3,4)$ code word $|\mathcal{R}_1|\mathcal{R}_2|$. However, letting $\mathcal{C}=|\mathcal{R}_1|\mathcal{R}_2|$ gives a code that differs by only one code word and has a threshold that is the same as the threshold of that best code up to four decimal places. (For the best $(3,6)$ code, $\mathcal{C}$ is equal to all the two-bit strings, which is the only possibility since there are only four two-bit strings.)
	
	For each number of code words $m$, the best codes appear to form a sequence in which the block length is increased by one by adding some fixed column to the code. This is clear particularly for $m=3$ and $m=4$: for $m=3$, adding the column $\scalebox{0.6}{$\begin{bmatrix}0\\1\\1\end{bmatrix}$}$ to the best $(n,3)$ code produces the best $(n+1,3)$ code, while for $m=4$ adding the column $\scalebox{0.6}{$\begin{bmatrix}0\\1\\0\\1\end{bmatrix}$}$ produces the best four-code-word code of the next block length. Increasing the block length in this way also increases the threshold, though the increase diminishes with increasing block length. These results are consistent with our results on repetition codes, in which adding the column $\scalebox{0.6}{$\begin{bmatrix}0\\1\end{bmatrix}$}$ to $\mathcal{R}_n$ produces $\mathcal{R}_{n+1}$, which has a higher threshold than $\mathcal{R}_n$. Just as the repetition code thresholds approach the Chau threshold in the limit $n\rightarrow\infty$, it appears from Fig. \ref{fig-plot_thresholds} that the sequence of best three-code-word codes approaches the same threshold as $n\rightarrow\infty$. This leads us to believe that the best codes for each number of code words will be a sequence of codes defined by successively adding some fixed column to the previous code in the sequence and that the thresholds of these codes will approach the Chau threshold.

\subsection{Directions for Future Work}\label{subsec-future_work}

	Based on our results from the previous sections, we now go through some directions for future work. First, finding the best codes for higher $(n,m)$ classes, particularly for the $(n,m)$ classes in Table \ref{table-inequivalent_sets} for which the best codes have not yet been found, will help in developing a better understanding of the structure of the best codes. As alluded to near the end of the previous section, another potential area for future work is examining the effect on the threshold of making very small changes to code words. See \cite{KhatriThesis} for examples of codes differing by only one code word that have very close thresholds. 
	
	Also at the end of the previous section, we made the observation that for the best codes examined in that section successively adding some fixed column to the best $(n,m)$ code gives the best $m$-code-word code for higher block lengths. To see if this observation might generalize to the best codes for all $m$, we are interested in examining the effect on the threshold of adding columns to a code. Specifically, for a given $(n,m)$ code $\mathcal{C}$, we are interested in the $2^m$ codes in the $(n+1,m)$ class obtained by adding each $m$-bit string as a column to $\mathcal{C}$. Note that the two codes obtained by adding a column and its complement  (the string obtained by flipping each bit) are equivalent. Furthermore, as we know from Sec. \ref{subsec-equivalent_codes}, adding a constant column to a code does not change the threshold. This leaves at most $2^{m-1}-1$ inequivalent codes without constant columns whose thresholds can be compared to that of $\mathcal{C}$ to see which, if any, increases the threshold.
	
	We are also interested in the following generalization of repetition codes that takes a $(n,m)$ code $\mathcal{C}$ and produces a $(n+1,2m)$ code defined as
	\begin{equation}\label{eq-complement_closed}
		 \left[\begin{tabular}{c|c}
			0 & \multirow{3}{1cm}{\centering$\mathcal{C}$}\\
			\shortstack{.\\.\\.} & \\
			0 & \\ \hline
			1 & \multirow{3}{1cm}{\centering$\overline{\mathcal{C}}$}\\
			\shortstack{.\\.\\.} & \\
			1 & \\ 
		\end{tabular}\right],
	\end{equation}
	where $\overline{\mathcal{C}}$ is called the \textit{complement} of $\mathcal{C}$ and is composed of the complement of each code word in $\mathcal{C}$. We call \eqref{eq-complement_closed} the \textit{complement extension} of $\mathcal{C}$. It is closed under taking the complement, in short \textit{complement-closed}, since each code word and its complement are contained in the code. Conversely, any complement-closed code is equivalent to the complement extension of some code. Complement-closed codes are a generalization of repetition codes that contain nonlinear codes. An even broader generalization of repetition codes can be obtained by replacing $\overline{\mathcal{C}}$ with $\pi(\mathcal{C})\oplus\alpha$, that is, with some code equivalent to $\mathcal{C}$. This extension, which we call the \textit{$(\pi,\alpha)$ extension}, contains the complement extension as a special case when $\pi$ is the identity permutation and $\alpha=11\dotsb 11$. We are interested in whether the $(\pi,\alpha)$ extension can increase the threshold of the original code $\mathcal{C}$, particularly when $\mathcal{C}$ is the best code in a class containing an odd number of code words. The reason for this is that for the best three-code-word codes \eqref{eq-best_three_codewords} there exist multiple $(\pi,\alpha)$ extensions that increase the threshold. In fact, one of them is always the complement extension, though it does not always provide the highest increase in the threshold. If we can thus show that for all the best codes with an odd number of code words there exists a $(\pi,\alpha)$ extension with a higher threshold, then we could restrict the search for codes exceeding the Chau threshold to those containing only an even number of code words.

\section{Postselection by Alice}\label{sec-Alice_postselection}

	As mentioned in the previous section, one of the challenges in testing codes with larger block lengths is that the dimension of Alice's space in the updated states \eqref{eq-filtered_state} is $2^n$. This results in SDPs that take a very long time to complete and often exceed the computer's memory limits. One way to address these issues is to allow Alice to postselect on the same code as Bob, resulting in the updated states
	\begin{equation}\label{eq-filtered_state_Alice}
		\begin{aligned}
		&(K_{\mathcal{C}}\otimes K_{\mathcal{C}})(\rho_Q^{AB})^{\otimes n}(K_{\mathcal{C}}\otimes K_{\mathcal{C}})^\dagger\\
		&\qquad=(K_{\mathcal{C}}\otimes\mathbbm{1}_{\tilde{B}})\rho_{Q,\mathcal{C}}^{A^n\tilde{B}}(K_{\mathcal{C}}\otimes\mathbbm{1}_{\tilde{B}})^\dagger.
		\end{aligned}
	\end{equation}
	This reduces the dimension of Alice's space to the same as Bob's and generally speeds up the SDP; however, it will not help to increase the threshold since the state \eqref{eq-filtered_state_Alice} is symmetrically extendable whenever the state $\rho_{Q,\mathcal{C}}^{A^n\tilde{B}}$ is symmetrically extendable. Since the converse of this statement is not necessarily true, postselection by Alice might \textit{decrease} the threshold. For the repetition codes, however, we find that this is not the case, that is, the thresholds with and without postselection by Alice are the same for all block lengths (see Appendix \ref{appendix-repetition_code_threshold} for a sketch of the proof). We have evidence to believe that the same might be true for the vast majority of codes, including the best codes from Sec. \ref{subsec-best_codes}, though we have found codes for which the thresholds are clearly different; see Appendix \ref{appendix-thresholds_Alice_post-selection} for some examples.
	
	In this section, we first consider a class of codes generalizing the repetition codes for which we were able to analytically determine the thresholds under postselection by Alice. We then describe a procedure to attempt a construction of a symmetric extension of any positive semidefinite operator that potentially avoids running the SDP \eqref{eq-symext_SDP}. By randomly selecting codes with larger block lengths and applying this procedure to the states \eqref{eq-filtered_state_Alice}, we then examine symmetric extendability in the gap in the case of Alice postselecting on the same code as Bob.
	
\subsection{Simplex Codes}\label{sec-simplex_codes}

	One natural way of generalizing repetition codes to more than two code words is to consider codes in which all pairs of distinct code words are some fixed Hamming distance, say $d$, away from each other. Such codes are called \textit{simplex codes} \cite{SloneMacWilliams}. Unlike the linear repetition codes, however, simplex codes can be nonlinear, which makes them of interest to us. Simplex codes $\mathcal{S}_{n,m,d}$ can be specified uniquely (up to equivalence and without constant columns) by three parameters: the block length $n$, the number of code words, $m$, and the constant Hamming distance $d$ separating the code words. When Alice and Bob postselect on the same simplex code, the state \eqref{eq-filtered_state_Alice}, that is,
	\begin{equation}\label{eq-filtered_state_Alice_simplex_code}
		(K_{\mathcal{S}_{n,m,d}}\otimes K_{\mathcal{S}_{n,m,d}})(\rho_Q^{AB})^{\otimes n}(K_{\mathcal{S}_{n,m,d}}\otimes K_{\mathcal{S}_{n,m,d}})^\dagger,
	\end{equation}
	can be shown to be diagonal in the $m$-dimensional Bell basis $\{\ket{\Phi_{k,\ell}}\}_{a,b=0}^{m-1}$ defined by $\ket{\Phi_{k,\ell}}=(\mathbbm{1}_{\mathbbm{C}^m}\otimes X(k)Z(\ell))\frac{1}{\sqrt{m}}\sum_{j=0}^{m-1}\ket{j,j}$, where $X(k)$, $Z(\ell)$ are the generalized Pauli (also called discrete Weyl) operators \cite{WatrousBook}. The state has only three distinct eigenvalues and is of the form
	\begin{equation}\label{eq-symext_Ranade}	
		\begin{aligned}
		\rho^{AB}&=\sum_{k,\ell=0}^{m-1}x_{k,\ell}\ket{\Phi_{k,\ell}}\bra{\Phi_{k,\ell}},\\
		x_{k,\ell}&=\left\{\begin{array}{c l} a & \text{if }k=\ell=0,\\b & \text{if }k=0,~\ell\geq 1,\\\frac{1-a-(m-1)b}{m(m-1)}&\text{otherwise},\end{array}\right.
		\end{aligned}
	\end{equation}
	where $a\geq b$ and $x\coloneqq a+(m-1)b\leq 1$. It is known \cite[Theorem 3]{RanadeSymExt} that such states are symmetrically extendable for all $m\geq 2$ if and only if
	\begin{equation}\label{eq-simplex_symext}
		a-b\leq 2\sqrt{\frac{(1-x)(2x-1)}{m-1}}+\frac{m-2}{m-1}(1-x).
	\end{equation}
	
	By determining the eigenvalues of the state \eqref{eq-filtered_state_Alice_simplex_code} and substituting them into Eq. \eqref{eq-simplex_symext}, the resulting condition at equality gives the symmetric extendability threshold of the state \eqref{eq-filtered_state_Alice_simplex_code}. For any simplex code, we find (see \cite{KhatriThesis} for details) that the threshold is never greater than the repetition code threshold for the same block length. In fact, as observed with the highest thresholds in Fig. \ref{fig-plot_thresholds}, for each block length the thresholds decrease as the number of code words increases beyond two, which means that nonlinear simplex codes are not better than the repetition codes for fixed block length. Furthermore, as the distance $d$ increases, we find that the thresholds approach the repetition code thresholds. In fact, in the limit $d\rightarrow\infty$, the thresholds approach $\frac{5-\sqrt{5}}{10}$, the same as the repetition code thresholds. This means that simplex codes cannot break symmetric extendability in the gap, at least in the case of Alice postselecting on the same code as Bob. This result also shows us that the repetition codes are not the only ones that can achieve the Chau threshold. While for repetition codes we were able to prove that the thresholds with and without postselection by Alice are the same for all block lengths, the method of proof used in that case did not work for simplex codes in general. Nevertheless, for some small simplex codes, we saw a difference between the thresholds on the order of $10^{-6}$ or less, which suggests that the thresholds might be the same.

\subsection{Constructing a Symmetric Extension and Random Search Over Larger Codes}\label{subsec-random_search}
	
	Though considering postselection by Alice on the same code as Bob generally reduces the runtime of the SDP \eqref{eq-symext_SDP}, to speed up the determination of symmetric extendability even further we developed a procedure to attempt a construction of a symmetric extension of the updated states \eqref{eq-filtered_state_Alice}. This procedure can also be used on the updated states \eqref{eq-filtered_state} without postselection by Alice (in fact, it can be used on any positive semidefinite operator), and is based on the following facts: every positive semidefinite operator $P^{AB}$ is uniquely associated with a completely positive (CP) map $\Phi$ from $A$ to $B$ (this is the Choi-Jamiolskowski correspondence) \cite{WatrousBook}; $P^{AB}$ is symmetrically extendable if and only if $\Phi$ is \textit{antidegradable}, that is, if and only if there exists a \textit{degrading channel} $\mathcal{E}$ such that $\mathcal{E}\circ\Phi^c=\Phi$, where $\Phi^c$ is a CP map complementary to $\Phi$ \cite{MyhrThesis}; if $P^{AB}$ is symmetrically extendable, then $P^{ABB'}\coloneqq(\mathbbm{1}_{AB}\otimes\mathcal{E})(\ket{\psi}\bra{\psi}^{ABE})$ is a symmetric extension, where $\ket{\psi}^{ABE}$ is a purification of $P^{AB}$ corresponding to $\Phi^c$ and $\mathcal{E}$ is a degrading channel from the definition of anti-degradability of $\Phi$ \cite{KhatriThesis}. In the procedure, we assume that a fixed orthonormal product basis $\{\ket{e_i}^A\otimes\ket{f_j}^B\equiv\ket{e_i,f_j}^{AB}\}_{i,j}$ has been chosen in which to represent $P^{AB}$. The procedure is as follows. 
	\begin{enumerate}
		\item Take $\ket{\psi}^{ABE_1E_2}=\text{vec}(\sqrt{P^{AB}})$ as a purification of $P^{AB}$, where the operation $\text{vec}$ is defined in the basis $\{\ket{e_i,f_j}^{AB}\}_{i,j}$ as $\text{vec}(\ket{e_i,f_j}\bra{e_{i'},f_{j'}})=\ket{e_i,f_j}\otimes\ket{e_{i'},f_{j'}}$. This operation defines the purification space $E$ as the tensor product $E_1\otimes E_2$, where $E_1$ is spanned by $\{\ket{e_i}\}_i$ and has the same dimension as $A$ and where $E_2$ is spanned by $\{\ket{f_j}\}_j$ and has the same dimension as $B$. 
		\item Use as ansatz a degrading map of the form $\mathcal{E}=\mathcal{N}\circ\Tr_{E_2}$, where the map $\mathcal{N}$ is defined by the condition that $\mathcal{E}$ satisfies the definition of antidegradability of $\Phi$.
		\item Since $\mathcal{N}$ must also be a channel, check that it is CP and trace preserving, which can be done, for example, using the Choi representation; see \cite{KhatriThesis} for the details. If it is, then $P^{AB}$ is symmetrically extendable; otherwise, run the SDP \eqref{eq-symext_SDP}.
	\end{enumerate}
	With this procedure, we can potentially avoid running the SDP \eqref{eq-symext_SDP}, thereby dramatically decreasing the amount of time it takes to determine symmetric extendability since determining whether a map is CP and trace preserving using the Choi representation can be done in much less time than an SDP.
	
	When applied to our problem at hand, this procedure allowed us to test over 540,000 codes in a reasonable amount of time. Specifically, we randomly selected 5000 codes (without constant columns) in all $(n,m)$ classes up to $(20,10)$, as well as the $(20,11)$, $(20,12)$, $(20,13)$, and $(20,14)$ classes, for a total of 548,818 codes \footnote{Not all of the codes selected initially were inequivalent. The total number of tested codes given is the number of inequivalent codes based on determining the equivalence of codes according to the results of the procedure described above. See \cite{KhatriThesis} for details.}. For each code, we tested symmetric extendability of the state \eqref{eq-filtered_state_Alice} within the gap at $Q=0.28,0.29,0.30,0.31,0.32$, and $0.33$. For approximately 99\% of the codes, the special map $\mathcal{E}$ from our procedure gave us a symmetric extension at all QBERs tested. For the remaining codes, the SDP \eqref{eq-symext_SDP} confirmed symmetric extendability for all QBERs tested. In other words, none of the selected codes could break symmetric extendability in the gap.

\section{Summary \& Outlook}

	We have examined two-way classical postprocessing protocols for PM-based six-state QKD protocols. Specifically, we were interested in determining the existence of such protocols in the gap of Fig. \ref{fig-1} for a particular class of states corresponding to Alice and Bob's measurement data. We used the results from Myhr et al. \cite{MyhrPaper} to reduce the problem to breaking symmetric extendability and showed that it is sufficient to consider postselection by Bob on error correction codes. Using this framework, we showed that repetition codes achieve the current-best Chau threshold that was previously obtained using the advantage distillation protocol, and we performed an exhaustive search over inequivalent codes of small block lengths and number of code words and observed that the repetition code has the highest updated threshold for each block length. In particular, for fixed block length, we found that increasing the number of code words tends to decrease the threshold. We then considered the less general case of postselection by Alice on the same code as Bob. We found that the simplex codes, a generalization of repetition codes to more than two code words, also achieve the Chau threshold. As well, from a random search on over 540,000 codes of larger block lengths and number of code words, we found that none of the codes broke symmetric extendability in the gap. Notably, our special map $\mathcal{E}$ was able to construct a symmetric extension throughout the gap for approximately 99\% of the codes tested.

	Based on our results, we conjecture that repetition codes are optimal for each block length and therefore that the Chau threshold is an upper bound on PM-based six-state QKD protocols with two-way classical postprocessing. This is in contrast to the entanglement-based version of such protocols in which a secret key can be distilled right up to the entanglement limit of $Q=\frac{1}{3}$ by allowing Alice and Bob to first perform \textit{quantum} postprocessing (e.g., entanglement distillation, quantum privacy amplification \cite{EPP_1,EPP_2,QPA_1,QPA_2}) before measuring their systems. As mentioned in the introduction, if repetition codes are indeed optimal, then it would also settle the question of the existence of the separation between secrecy formation and secrecy extraction and of the existence of bound secrecy. We thus hope that our results can guide any future attempts at a proof of the optimality of repetition codes.
	
\begin{acknowledgments}
	
	We thank the NSERC Discovery Grant and Industry Canada for financial support. SK also acknowledges financial support from the NSERC Canada Graduate Scholarship and the Ontario Graduate Scholarship.

\end{acknowledgments}

\bibliography{references}{}

\begin{thebibliography}{33}%
\makeatletter
\providecommand \@ifxundefined [1]{%
 \@ifx{#1\undefined}
}%
\providecommand \@ifnum [1]{%
 \ifnum #1\expandafter \@firstoftwo
 \else \expandafter \@secondoftwo
 \fi
}%
\providecommand \@ifx [1]{%
 \ifx #1\expandafter \@firstoftwo
 \else \expandafter \@secondoftwo
 \fi
}%
\providecommand \natexlab [1]{#1}%
\providecommand \enquote  [1]{``#1''}%
\providecommand \bibnamefont  [1]{#1}%
\providecommand \bibfnamefont [1]{#1}%
\providecommand \citenamefont [1]{#1}%
\providecommand \href@noop [0]{\@secondoftwo}%
\providecommand \href [0]{\begingroup \@sanitize@url \@href}%
\providecommand \@href[1]{\@@startlink{#1}\@@href}%
\providecommand \@@href[1]{\endgroup#1\@@endlink}%
\providecommand \@sanitize@url [0]{\catcode `\\12\catcode `\$12\catcode
  `\&12\catcode `\#12\catcode `\^12\catcode `\_12\catcode `\%12\relax}%
\providecommand \@@startlink[1]{}%
\providecommand \@@endlink[0]{}%
\providecommand \url  [0]{\begingroup\@sanitize@url \@url }%
\providecommand \@url [1]{\endgroup\@href {#1}{\urlprefix }}%
\providecommand \urlprefix  [0]{URL }%
\providecommand \Eprint [0]{\href }%
\providecommand \doibase [0]{http://dx.doi.org/}%
\providecommand \selectlanguage [0]{\@gobble}%
\providecommand \bibinfo  [0]{\@secondoftwo}%
\providecommand \bibfield  [0]{\@secondoftwo}%
\providecommand \translation [1]{[#1]}%
\providecommand \BibitemOpen [0]{}%
\providecommand \bibitemStop [0]{}%
\providecommand \bibitemNoStop [0]{.\EOS\space}%
\providecommand \EOS [0]{\spacefactor3000\relax}%
\providecommand \BibitemShut  [1]{\csname bibitem#1\endcsname}%
\let\auto@bib@innerbib\@empty
\bibitem [{\citenamefont {Curty}\ \emph {et~al.}(2004)\citenamefont {Curty},
  \citenamefont {Lewenstein},\ and\ \citenamefont
  {L\"utkenhaus}}]{CurtyEntanglement}%
  \BibitemOpen
  \bibfield  {author} {\bibinfo {author} {\bibfnamefont {M.}~\bibnamefont
  {Curty}}, \bibinfo {author} {\bibfnamefont {M.}~\bibnamefont {Lewenstein}}, \
  and\ \bibinfo {author} {\bibfnamefont {N.}~\bibnamefont {L\"utkenhaus}},\
  }\href {\doibase 10.1103/PhysRevLett.92.217903} {\bibfield  {journal}
  {\bibinfo  {journal} {Phys. Rev. Lett.}\ }\textbf {\bibinfo {volume} {92}},\
  \bibinfo {pages} {217903} (\bibinfo {year} {2004})}\BibitemShut {NoStop}%
\bibitem [{Note1()}]{Note1}%
  \BibitemOpen
  \bibinfo {note} {See \cite {MyhrThesis} for an introduction to symmetrically
  extendable states. See \cite {ChenSymext,RanadeSymExt} for the only
  currently-known necessary and sufficient criteria for symmetric extendability
  that hold for two classes of states. From now on, by symmetrically extendable
  we will always mean symmetrically extendable to a copy of Bob's
  system.}\BibitemShut {Stop}%
\bibitem [{\citenamefont {Moroder}\ \emph {et~al.}(2006)\citenamefont
  {Moroder}, \citenamefont {Curty},\ and\ \citenamefont
  {L\"utkenhaus}}]{MoroderOneWay}%
  \BibitemOpen
  \bibfield  {author} {\bibinfo {author} {\bibfnamefont {T.}~\bibnamefont
  {Moroder}}, \bibinfo {author} {\bibfnamefont {M.}~\bibnamefont {Curty}}, \
  and\ \bibinfo {author} {\bibfnamefont {N.}~\bibnamefont {L\"utkenhaus}},\
  }\href {\doibase 10.1103/PhysRevA.74.052301} {\bibfield  {journal} {\bibinfo
  {journal} {Phys. Rev. A}\ }\textbf {\bibinfo {volume} {74}},\ \bibinfo
  {pages} {052301} (\bibinfo {year} {2006})}\BibitemShut {NoStop}%
\bibitem [{\citenamefont {Bru\ss{}}(1998)}]{BrussSix-State}%
  \BibitemOpen
  \bibfield  {author} {\bibinfo {author} {\bibfnamefont {D.}~\bibnamefont
  {Bru\ss{}}},\ }\href {\doibase 10.1103/PhysRevLett.81.3018} {\bibfield
  {journal} {\bibinfo  {journal} {Phys. Rev. Lett.}\ }\textbf {\bibinfo
  {volume} {81}},\ \bibinfo {pages} {3018} (\bibinfo {year}
  {1998})}\BibitemShut {NoStop}%
\bibitem [{\citenamefont {Myhr}(2010)}]{MyhrThesis}%
  \BibitemOpen
  \bibfield  {author} {\bibinfo {author} {\bibfnamefont {G.~O.}\ \bibnamefont
  {Myhr}},\ }\emph {\bibinfo {title} {Symmetric Extension of Bipartite Quantum
  States and its Use in Quantum Key Distribution with Two-Way
  Postprocessing}},\ \href {https://arxiv.org/abs/1103.0766} {Ph.D. thesis},\
  \bibinfo  {school} {Friedrich-Alexander-Universit\"{a}t
  Erlangen-N\"{u}remberg} (\bibinfo {year} {2010})\BibitemShut {NoStop}%
\bibitem [{\citenamefont {Gottesman}\ and\ \citenamefont
  {Lo}(2003)}]{GottesmanLo}%
  \BibitemOpen
  \bibfield  {author} {\bibinfo {author} {\bibfnamefont {D.}~\bibnamefont
  {Gottesman}}\ and\ \bibinfo {author} {\bibfnamefont {H.-K.}\ \bibnamefont
  {Lo}},\ }\href@noop {} {\bibfield  {journal} {\bibinfo  {journal} {IEEE
  Transactions on Information Theory}\ }\textbf {\bibinfo {volume} {49}},\
  \bibinfo {pages} {457} (\bibinfo {year} {2003})}\BibitemShut {NoStop}%
\bibitem [{\citenamefont {Chau}(2002)}]{Chau}%
  \BibitemOpen
  \bibfield  {author} {\bibinfo {author} {\bibfnamefont {H.~F.}\ \bibnamefont
  {Chau}},\ }\href {\doibase 10.1103/PhysRevA.66.060302} {\bibfield  {journal}
  {\bibinfo  {journal} {Phys. Rev. A}\ }\textbf {\bibinfo {volume} {66}},\
  \bibinfo {pages} {060302} (\bibinfo {year} {2002})}\BibitemShut {NoStop}%
\bibitem [{\citenamefont {Bennett}\ and\ \citenamefont
  {Brassard}(1984)}]{BB84}%
  \BibitemOpen
  \bibfield  {author} {\bibinfo {author} {\bibfnamefont {C.~H.}\ \bibnamefont
  {Bennett}}\ and\ \bibinfo {author} {\bibfnamefont {G.}~\bibnamefont
  {Brassard}},\ }in\ \href@noop {} {\emph {\bibinfo {booktitle} {Proceedings of
  IEEE International Conference on Computers, Systems, and Signal Processing,
  Bangalore, India}}}\ (\bibinfo {year} {1984})\ pp.\ \bibinfo {pages}
  {175--179}\BibitemShut {NoStop}%
\bibitem [{\citenamefont {Shor}\ and\ \citenamefont
  {Preskill}(2000)}]{ShorPreskill_BB84}%
  \BibitemOpen
  \bibfield  {author} {\bibinfo {author} {\bibfnamefont {P.~W.}\ \bibnamefont
  {Shor}}\ and\ \bibinfo {author} {\bibfnamefont {J.}~\bibnamefont
  {Preskill}},\ }\href {\doibase 10.1103/PhysRevLett.85.441} {\bibfield
  {journal} {\bibinfo  {journal} {Phys. Rev. Lett.}\ }\textbf {\bibinfo
  {volume} {85}},\ \bibinfo {pages} {441} (\bibinfo {year} {2000})}\BibitemShut
  {NoStop}%
\bibitem [{\citenamefont {Maurer}(1993)}]{MaurerAD}%
  \BibitemOpen
  \bibfield  {author} {\bibinfo {author} {\bibfnamefont {U.~M.}\ \bibnamefont
  {Maurer}},\ }\href@noop {} {\bibfield  {journal} {\bibinfo  {journal} {IEEE
  Transactions on Information Theory}\ }\textbf {\bibinfo {volume} {39}},\
  \bibinfo {pages} {733} (\bibinfo {year} {1993})}\BibitemShut {NoStop}%
\bibitem [{\citenamefont {Ac\'{\i}n}\ \emph {et~al.}(2006)\citenamefont
  {Ac\'{\i}n}, \citenamefont {Bae}, \citenamefont {Bagan}, \citenamefont
  {Baig}, \citenamefont {Masanes},\ and\ \citenamefont {Mu\~noz
  Tapia}}]{Acin1}%
  \BibitemOpen
  \bibfield  {author} {\bibinfo {author} {\bibfnamefont {A.}~\bibnamefont
  {Ac\'{\i}n}}, \bibinfo {author} {\bibfnamefont {J.}~\bibnamefont {Bae}},
  \bibinfo {author} {\bibfnamefont {E.}~\bibnamefont {Bagan}}, \bibinfo
  {author} {\bibfnamefont {M.}~\bibnamefont {Baig}}, \bibinfo {author}
  {\bibfnamefont {L.}~\bibnamefont {Masanes}}, \ and\ \bibinfo {author}
  {\bibfnamefont {R.}~\bibnamefont {Mu\~noz Tapia}},\ }\href {\doibase
  10.1103/PhysRevA.73.012327} {\bibfield  {journal} {\bibinfo  {journal} {Phys.
  Rev. A}\ }\textbf {\bibinfo {volume} {73}},\ \bibinfo {pages} {012327}
  (\bibinfo {year} {2006})}\BibitemShut {NoStop}%
\bibitem [{\citenamefont {Bae}\ and\ \citenamefont
  {Ac\'{\i}n}(2007)}]{BaeAcin}%
  \BibitemOpen
  \bibfield  {author} {\bibinfo {author} {\bibfnamefont {J.}~\bibnamefont
  {Bae}}\ and\ \bibinfo {author} {\bibfnamefont {A.}~\bibnamefont
  {Ac\'{\i}n}},\ }\href {\doibase 10.1103/PhysRevA.75.012334} {\bibfield
  {journal} {\bibinfo  {journal} {Phys. Rev. A}\ }\textbf {\bibinfo {volume}
  {75}},\ \bibinfo {pages} {012334} (\bibinfo {year} {2007})}\BibitemShut
  {NoStop}%
\bibitem [{\citenamefont {Myhr}\ \emph {et~al.}(2009)\citenamefont {Myhr},
  \citenamefont {Renes}, \citenamefont {Doherty},\ and\ \citenamefont
  {L\"utkenhaus}}]{MyhrPaper}%
  \BibitemOpen
  \bibfield  {author} {\bibinfo {author} {\bibfnamefont {G.~O.}\ \bibnamefont
  {Myhr}}, \bibinfo {author} {\bibfnamefont {J.~M.}\ \bibnamefont {Renes}},
  \bibinfo {author} {\bibfnamefont {A.~C.}\ \bibnamefont {Doherty}}, \ and\
  \bibinfo {author} {\bibfnamefont {N.}~\bibnamefont {L\"utkenhaus}},\ }\href
  {\doibase 10.1103/PhysRevA.79.042329} {\bibfield  {journal} {\bibinfo
  {journal} {Phys. Rev. A}\ }\textbf {\bibinfo {volume} {79}},\ \bibinfo
  {pages} {042329} (\bibinfo {year} {2009})}\BibitemShut {NoStop}%
\bibitem [{\citenamefont {Ac\'{\i}n}\ and\ \citenamefont
  {Gisin}(2005)}]{Acin3}%
  \BibitemOpen
  \bibfield  {author} {\bibinfo {author} {\bibfnamefont {A.}~\bibnamefont
  {Ac\'{\i}n}}\ and\ \bibinfo {author} {\bibfnamefont {N.}~\bibnamefont
  {Gisin}},\ }\href {\doibase 10.1103/PhysRevLett.94.020501} {\bibfield
  {journal} {\bibinfo  {journal} {Phys. Rev. Lett.}\ }\textbf {\bibinfo
  {volume} {94}},\ \bibinfo {pages} {020501} (\bibinfo {year}
  {2005})}\BibitemShut {NoStop}%
\bibitem [{\citenamefont {Gisin}\ and\ \citenamefont {Wolf}(2000)}]{Gisin2000}%
  \BibitemOpen
  \bibfield  {author} {\bibinfo {author} {\bibfnamefont {N.}~\bibnamefont
  {Gisin}}\ and\ \bibinfo {author} {\bibfnamefont {S.}~\bibnamefont {Wolf}},\
  }\enquote {\bibinfo {title} {Linking classical and quantum key agreement: Is
  there ``bound information''?}}\ in\ \href {\doibase 10.1007/3-540-44598-6_30}
  {\emph {\bibinfo {booktitle} {Advances in Cryptology --- CRYPTO 2000: 20th
  Annual International Cryptology Conference Santa Barbara, California, USA,
  August 20--24, 2000 Proceedings}}},\ \bibinfo {editor} {edited by\ \bibinfo
  {editor} {\bibfnamefont {M.}~\bibnamefont {Bellare}}}\ (\bibinfo  {publisher}
  {Springer Berlin Heidelberg},\ \bibinfo {address} {Berlin, Heidelberg},\
  \bibinfo {year} {2000})\ pp.\ \bibinfo {pages} {482--500}\BibitemShut
  {NoStop}%
\bibitem [{\citenamefont {Gisin}\ \emph {et~al.}(2001)\citenamefont {Gisin},
  \citenamefont {Renner},\ and\ \citenamefont {Wolf}}]{Gisin2001}%
  \BibitemOpen
  \bibfield  {author} {\bibinfo {author} {\bibfnamefont {N.}~\bibnamefont
  {Gisin}}, \bibinfo {author} {\bibfnamefont {R.}~\bibnamefont {Renner}}, \
  and\ \bibinfo {author} {\bibfnamefont {S.}~\bibnamefont {Wolf}},\ }\enquote
  {\bibinfo {title} {Bound information: The classical analog to bound quantum
  entanglemen},}\ in\ \href {\doibase 10.1007/978-3-0348-8266-8_38} {\emph
  {\bibinfo {booktitle} {European Congress of Mathematics: Barcelona, July
  10--14, 2000 Volume II}}},\ \bibinfo {editor} {edited by\ \bibinfo {editor}
  {\bibfnamefont {C.}~\bibnamefont {Casacuberta}}, \bibinfo {editor}
  {\bibfnamefont {R.~M.}\ \bibnamefont {Mir{\'o}-Roig}}, \bibinfo {editor}
  {\bibfnamefont {J.}~\bibnamefont {Verdera}}, \ and\ \bibinfo {editor}
  {\bibfnamefont {S.}~\bibnamefont {Xamb{\'o}-Descamps}}}\ (\bibinfo
  {publisher} {Birkh{\"a}user Basel},\ \bibinfo {address} {Basel},\ \bibinfo
  {year} {2001})\ pp.\ \bibinfo {pages} {439--447}\BibitemShut {NoStop}%
\bibitem [{\citenamefont {Gisin}\ \emph {et~al.}(2002)\citenamefont {Gisin},
  \citenamefont {Renner},\ and\ \citenamefont {Wolf}}]{Gisin2002}%
  \BibitemOpen
  \bibfield  {author} {\bibinfo {author} {\bibnamefont {Gisin}}, \bibinfo
  {author} {\bibnamefont {Renner}}, \ and\ \bibinfo {author} {\bibnamefont
  {Wolf}},\ }\href {\doibase 10.1007/s00453-002-0972-7} {\bibfield  {journal}
  {\bibinfo  {journal} {Algorithmica}\ }\textbf {\bibinfo {volume} {34}},\
  \bibinfo {pages} {389} (\bibinfo {year} {2002})}\BibitemShut {NoStop}%
\bibitem [{\citenamefont {Collins}\ and\ \citenamefont
  {Popescu}(2002)}]{Entanglement_classical_analog}%
  \BibitemOpen
  \bibfield  {author} {\bibinfo {author} {\bibfnamefont {D.}~\bibnamefont
  {Collins}}\ and\ \bibinfo {author} {\bibfnamefont {S.}~\bibnamefont
  {Popescu}},\ }\href {\doibase 10.1103/PhysRevA.65.032321} {\bibfield
  {journal} {\bibinfo  {journal} {Phys. Rev. A}\ }\textbf {\bibinfo {volume}
  {65}},\ \bibinfo {pages} {032321} (\bibinfo {year} {2002})}\BibitemShut
  {NoStop}%
\bibitem [{\citenamefont {Renner}\ and\ \citenamefont {Wolf}(2003)}]{RenWol03}%
  \BibitemOpen
  \bibfield  {author} {\bibinfo {author} {\bibfnamefont {R.}~\bibnamefont
  {Renner}}\ and\ \bibinfo {author} {\bibfnamefont {S.}~\bibnamefont {Wolf}},\
  }in\ \href@noop {} {\emph {\bibinfo {booktitle} {Advances in Cryptology ---
  EUROCRYPT 2003}}},\ \bibinfo {series} {Lecture Notes in Computer Science},
  Vol.\ \bibinfo {volume} {2656},\ \bibinfo {editor} {edited by\ \bibinfo
  {editor} {\bibfnamefont {E.}~\bibnamefont {Biham}}}\ (\bibinfo  {publisher}
  {Springer-Verlag},\ \bibinfo {year} {2003})\ pp.\ \bibinfo {pages}
  {562--577}\BibitemShut {NoStop}%
\bibitem [{\citenamefont {L\"{u}tkenhaus}(2014)}]{LutkenhausChapter}%
  \BibitemOpen
  \bibfield  {author} {\bibinfo {author} {\bibfnamefont {N.}~\bibnamefont
  {L\"{u}tkenhaus}},\ }in\ \href@noop {} {\emph {\bibinfo {booktitle} {Quantum
  Information and Coherence}}},\ \bibinfo {editor} {edited by\ \bibinfo
  {editor} {\bibfnamefont {E.}~\bibnamefont {Andersson}}\ and\ \bibinfo
  {editor} {\bibfnamefont {P.}~\bibnamefont {\"{O}hberg}}}\ (\bibinfo
  {publisher} {Springer},\ \bibinfo {year} {2014})\ Chap.~\bibinfo {chapter}
  {10}, pp.\ \bibinfo {pages} {107--146}\BibitemShut {NoStop}%
\bibitem [{Note2()}]{Note2}%
  \BibitemOpen
  \bibinfo {note} {Normalization of the state is unimportant for symmetric
  extendability since if $\rho ^{AB}$ is symmetrically extendable then so is
  $\alpha \rho ^{AB}$ for any $\alpha >0$.}\BibitemShut {Stop}%
\bibitem [{\citenamefont {Khatri}(2016)}]{KhatriThesis}%
  \BibitemOpen
  \bibfield  {author} {\bibinfo {author} {\bibfnamefont {S.}~\bibnamefont
  {Khatri}},\ }\emph {\bibinfo {title} {Symmetric Extendability of Quantum
  States and the Extreme Limits of Quantum Key Distribution}},\ \href
  {https://uwspace.uwaterloo.ca/handle/10012/10993} {Master's thesis},\
  \bibinfo  {school} {University of Waterloo} (\bibinfo {year}
  {2016})\BibitemShut {NoStop}%
\bibitem [{\citenamefont {MacWilliams}\ and\ \citenamefont
  {Sloane}(1977)}]{SloneMacWilliams}%
  \BibitemOpen
  \bibfield  {author} {\bibinfo {author} {\bibfnamefont {F.~J.}\ \bibnamefont
  {MacWilliams}}\ and\ \bibinfo {author} {\bibfnamefont {N.~J.~A.}\
  \bibnamefont {Sloane}},\ }\href@noop {} {\emph {\bibinfo {title} {The Theory
  of Error-Correcting Codes}}}\ (\bibinfo  {publisher} {North-Holland
  Publishing Company},\ \bibinfo {year} {1977})\BibitemShut {NoStop}%
\bibitem [{\citenamefont {Lofberg}(2004)}]{yalmip}%
  \BibitemOpen
  \bibfield  {author} {\bibinfo {author} {\bibfnamefont {J.}~\bibnamefont
  {Lofberg}},\ }in\ \href {\doibase 10.1109/CACSD.2004.1393890} {\emph
  {\bibinfo {booktitle} {Computer Aided Control Systems Design, 2004 IEEE
  International Symposium on}}}\ (\bibinfo {year} {2004})\ pp.\ \bibinfo
  {pages} {284--289}\BibitemShut {NoStop}%
\bibitem [{\citenamefont {O'Donoghue}\ \emph {et~al.}(2016)\citenamefont
  {O'Donoghue}, \citenamefont {Chu}, \citenamefont {Parikh},\ and\
  \citenamefont {Boyd}}]{SCS_solver}%
  \BibitemOpen
  \bibfield  {author} {\bibinfo {author} {\bibfnamefont {B.}~\bibnamefont
  {O'Donoghue}}, \bibinfo {author} {\bibfnamefont {E.}~\bibnamefont {Chu}},
  \bibinfo {author} {\bibfnamefont {N.}~\bibnamefont {Parikh}}, \ and\ \bibinfo
  {author} {\bibfnamefont {S.}~\bibnamefont {Boyd}},\ }\href {\doibase
  10.1007/s10957-016-0892-3} {\bibfield  {journal} {\bibinfo  {journal}
  {Journal of Optimization Theory and Applications}\ }\textbf {\bibinfo
  {volume} {169}},\ \bibinfo {pages} {1042} (\bibinfo {year}
  {2016})}\BibitemShut {NoStop}%
\bibitem [{\citenamefont {Watrous}(2016)}]{WatrousBook}%
  \BibitemOpen
  \bibfield  {author} {\bibinfo {author} {\bibfnamefont {J.}~\bibnamefont
  {Watrous}},\ }\href {https://cs.uwaterloo.ca/~watrous/TQI/} {\enquote
  {\bibinfo {title} {Theory of quantum information},}\ } (\bibinfo {year}
  {2016})\BibitemShut {NoStop}%
\bibitem [{\citenamefont {Ranade}(2009)}]{RanadeSymExt}%
  \BibitemOpen
  \bibfield  {author} {\bibinfo {author} {\bibfnamefont {K.~S.}\ \bibnamefont
  {Ranade}},\ }\href {http://stacks.iop.org/1751-8121/42/i=42/a=425302}
  {\bibfield  {journal} {\bibinfo  {journal} {Journal of Physics A:
  Mathematical and Theoretical}\ }\textbf {\bibinfo {volume} {42}},\ \bibinfo
  {pages} {425302} (\bibinfo {year} {2009})}\BibitemShut {NoStop}%
\bibitem [{Note3()}]{Note3}%
  \BibitemOpen
  \bibinfo {note} {Not all of the codes selected initially were inequivalent.
  The total number of tested codes given is the number of inequivalent codes
  based on determining the equivalence of codes according to the results of the
  procedure described above. See \cite {KhatriThesis} for details.}\BibitemShut
  {Stop}%
\bibitem [{\citenamefont {Bennett}\ \emph
  {et~al.}(1996{\natexlab{a}})\citenamefont {Bennett}, \citenamefont
  {Brassard}, \citenamefont {Popescu}, \citenamefont {Schumacher},
  \citenamefont {Smolin},\ and\ \citenamefont {Wootters}}]{EPP_1}%
  \BibitemOpen
  \bibfield  {author} {\bibinfo {author} {\bibfnamefont {C.~H.}\ \bibnamefont
  {Bennett}}, \bibinfo {author} {\bibfnamefont {G.}~\bibnamefont {Brassard}},
  \bibinfo {author} {\bibfnamefont {S.}~\bibnamefont {Popescu}}, \bibinfo
  {author} {\bibfnamefont {B.}~\bibnamefont {Schumacher}}, \bibinfo {author}
  {\bibfnamefont {J.~A.}\ \bibnamefont {Smolin}}, \ and\ \bibinfo {author}
  {\bibfnamefont {W.~K.}\ \bibnamefont {Wootters}},\ }\href {\doibase
  10.1103/PhysRevLett.76.722} {\bibfield  {journal} {\bibinfo  {journal} {Phys.
  Rev. Lett.}\ }\textbf {\bibinfo {volume} {76}},\ \bibinfo {pages} {722}
  (\bibinfo {year} {1996}{\natexlab{a}})}\BibitemShut {NoStop}%
\bibitem [{\citenamefont {Bennett}\ \emph
  {et~al.}(1996{\natexlab{b}})\citenamefont {Bennett}, \citenamefont
  {DiVincenzo}, \citenamefont {Smolin},\ and\ \citenamefont
  {Wootters}}]{EPP_2}%
  \BibitemOpen
  \bibfield  {author} {\bibinfo {author} {\bibfnamefont {C.~H.}\ \bibnamefont
  {Bennett}}, \bibinfo {author} {\bibfnamefont {D.~P.}\ \bibnamefont
  {DiVincenzo}}, \bibinfo {author} {\bibfnamefont {J.~A.}\ \bibnamefont
  {Smolin}}, \ and\ \bibinfo {author} {\bibfnamefont {W.~K.}\ \bibnamefont
  {Wootters}},\ }\href {\doibase 10.1103/PhysRevA.54.3824} {\bibfield
  {journal} {\bibinfo  {journal} {Phys. Rev. A}\ }\textbf {\bibinfo {volume}
  {54}},\ \bibinfo {pages} {3824} (\bibinfo {year}
  {1996}{\natexlab{b}})}\BibitemShut {NoStop}%
\bibitem [{\citenamefont {Bennett}\ and\ \citenamefont
  {Brassard}(1985)}]{QPA_1}%
  \BibitemOpen
  \bibfield  {author} {\bibinfo {author} {\bibfnamefont {C.}~\bibnamefont
  {Bennett}}\ and\ \bibinfo {author} {\bibfnamefont {G.}~\bibnamefont
  {Brassard}},\ }\href@noop {} {\bibfield  {journal} {\bibinfo  {journal} {IBM
  Technical Disclosure Bulletin}\ }\textbf {\bibinfo {volume} {28}},\ \bibinfo
  {pages} {3153} (\bibinfo {year} {1985})}\BibitemShut {NoStop}%
\bibitem [{\citenamefont {Deutsch}\ \emph {et~al.}(1996)\citenamefont
  {Deutsch}, \citenamefont {Ekert}, \citenamefont {Jozsa}, \citenamefont
  {Macchiavello}, \citenamefont {Popescu},\ and\ \citenamefont
  {Sanpera}}]{QPA_2}%
  \BibitemOpen
  \bibfield  {author} {\bibinfo {author} {\bibfnamefont {D.}~\bibnamefont
  {Deutsch}}, \bibinfo {author} {\bibfnamefont {A.}~\bibnamefont {Ekert}},
  \bibinfo {author} {\bibfnamefont {R.}~\bibnamefont {Jozsa}}, \bibinfo
  {author} {\bibfnamefont {C.}~\bibnamefont {Macchiavello}}, \bibinfo {author}
  {\bibfnamefont {S.}~\bibnamefont {Popescu}}, \ and\ \bibinfo {author}
  {\bibfnamefont {A.}~\bibnamefont {Sanpera}},\ }\href {\doibase
  10.1103/PhysRevLett.77.2818} {\bibfield  {journal} {\bibinfo  {journal}
  {Phys. Rev. Lett.}\ }\textbf {\bibinfo {volume} {77}},\ \bibinfo {pages}
  {2818} (\bibinfo {year} {1996})}\BibitemShut {NoStop}%
\bibitem [{\citenamefont {Chen}\ \emph {et~al.}(2014)\citenamefont {Chen},
  \citenamefont {Ji}, \citenamefont {Kribs}, \citenamefont {L\"utkenhaus},\
  and\ \citenamefont {Zeng}}]{ChenSymext}%
  \BibitemOpen
  \bibfield  {author} {\bibinfo {author} {\bibfnamefont {J.}~\bibnamefont
  {Chen}}, \bibinfo {author} {\bibfnamefont {Z.}~\bibnamefont {Ji}}, \bibinfo
  {author} {\bibfnamefont {D.}~\bibnamefont {Kribs}}, \bibinfo {author}
  {\bibfnamefont {N.}~\bibnamefont {L\"utkenhaus}}, \ and\ \bibinfo {author}
  {\bibfnamefont {B.}~\bibnamefont {Zeng}},\ }\href {\doibase
  10.1103/PhysRevA.90.032318} {\bibfield  {journal} {\bibinfo  {journal} {Phys.
  Rev. A}\ }\textbf {\bibinfo {volume} {90}},\ \bibinfo {pages} {032318}
  (\bibinfo {year} {2014})}\BibitemShut {NoStop}%
\end{thebibliography}%

\appendix

\section{Repetition Code Thresholds}\label{appendix-repetition_code_threshold}

	In this section, we provide a sketch of the proof that the solution to Eq. \eqref{eq-repetition_symext_equation} gives the repetition code thresholds $Q_{\mathcal{R}_n}^*$. This proof also establishes the fact that the thresholds with and without postselection by Alice on repetition codes are the same. For all the details, see \cite{KhatriThesis}.
	
	We start with the state in Eq. \eqref{eq-filtered_state_Alice} for the case when Alice and Bob postselect on the repetition code $\mathcal{R}_n$:
	\begin{equation}\label{eq-filtered_state_Alice_repetition_code}
		(K_{\mathcal{R}_n}\otimes K_{\mathcal{R}_n})(\rho_Q^{AB})^{\otimes n}(K_{\mathcal{R}_n}\otimes K_{\mathcal{R}_n})^\dagger.
	\end{equation}
	This is a two-qubit state, and it is known from \cite{ChenSymext} that any two-qubit state $\rho^{AB}$ is symmetrically extendable if and only if
	\begin{equation}\label{eq-two_qubit_symext}
		\Tr[(\rho^B)^2]\geq\Tr[(\rho^{AB})^2]-4\sqrt{\text{det}(\rho^{AB})},
	\end{equation}
	where $\rho^B=\Tr_A[\rho^{AB}]$. By substituting the state \eqref{eq-filtered_state_Alice_repetition_code} into \eqref{eq-two_qubit_symext} and simplifying, we obtain
	\begin{equation}\label{eq-repetition_symext_inequality}
		4Q^{2n}-4Q^n(1-Q)^n+(1-2Q)^{2n}\leq 0.
	\end{equation}
	At equality, this is Eq. \eqref{eq-repetition_symext_equation} and its solution gives the symmetric extendability thresholds of the state \eqref{eq-filtered_state_Alice_repetition_code} with postselection by Alice, while we are interested in the symmetric extendability thresholds of the state \eqref{eq-filtered_state} \textit{without} postselection by Alice. Remarkably, by applying the special procedure of Sec. \ref{subsec-random_search} to the state $\rho_{Q,\mathcal{R}_n}^{A^n\tilde{B}}$ without postselection by Alice, we get that the map $\mathcal{N}$ is completely positive and trace preserving whenever the condition \eqref{eq-repetition_symext_inequality} is satisfied. In other words, if the state \eqref{eq-filtered_state_Alice_repetition_code} is symmetrically extendable, then so is $\rho_{Q,\mathcal{R}_n}^{A^n\tilde{B}}$. Since the converse of this statement is true, it holds that $\rho_{Q,\mathcal{R}_n}^{A^n\tilde{B}}$ is symmetrically extendable if and only if \eqref{eq-filtered_state_Alice_repetition_code} is symmetrically extendable. This means that the solution to Eq. \eqref{eq-repetition_symext_equation} gives the threshold $Q_{\mathcal{R}_n}^*$ and therefore that the thresholds with and without postselection by Alice are the same for repetition codes. Whether there exist other classes of codes, in addition to the repetition codes, for which the thresholds are the same is an open problem.

\section{Thresholds With and Without Postselection by Alice}\label{appendix-thresholds_Alice_post-selection}

	In \cite{KhatriThesis}, we determined for each of the inequivalent codes tested in Sec. \ref{sec-thresholds} the corresponding threshold for the states \eqref{eq-filtered_state_Alice} with postselection by Alice on the same code as Bob. For the vast majority of these codes, the difference between the two thresholds was on the order of $10^{-4}$ or less. (Recall from Sec. \ref{subsec-numerics_thresholds} that our procedure for determining the threshold is accurate to within $8.5\times 10^{-4}$.)  For some codes, however, the difference was higher. Table \ref{tab-high_difference} shows some of the codes with the highest difference. The smallest difference is $1.5\times 10^{-3}$ and the largest difference is $6.8\times 10^{-3}$. These differences are only marginally greater than the $8.5\times 10^{-4}$ within which the thresholds themselves are accurate. A more refined analysis should therefore be performed to confirm the differences seen here. For example, one can take a small interval around the thresholds given in the table and perform the same procedure described in Sec. \ref{subsec-numerics_thresholds}. The number of points in the interval should be large enough so that the spacing between points is smaller than $8.5\times 10^{-4}$ and thus the accuracy of the threshold is higher.
	
	\begin{table}[h]
		\centering
		\begin{tabular}{|c|c||c|c|}
		\hline \multirow{2}{*}{$\left[\begin{array}{c}000\\100\\110\\111\end{array}\right]$}& \sr{-0.45cm}{1.1cm} 0.1965 & \multirow{2}{*}{$\left[\begin{array}{c} 0000\\1000\\1110\\1111\end{array}\right]$} & 0.2196 \\\cline{2-2}\cline{4-4}
		& \sr{-0.45cm}{1.1cm} 0.1948 & & 0.2154 \\ \hline\hline
		\multirow{2}{*}{$\left[\begin{array}{c}0000\\1100\\1110\\1111\end{array}\right]$}& \sr{-0.45cm}{1.1cm} 0.2158 & \multirow{2}{*}{$\left[\begin{array}{c}00000\\10000\\11110\\11111\end{array}\right]$} & 0.2330 \\\cline{2-2}\cline{4-4}
		& \sr{-0.45cm}{1.1cm} 0.2143 & & 0.2281 \\ \hline\hline
		\multirow{2}{*}[0.25cm]{$\left[\begin{array}{c}000\\100\\101\\110\\111\end{array}\right]$}& \sr{-0.45cm}{1.1cm} 0.1833 & \multirow{2}{*}[0.25cm]{$\left[\begin{array}{c}0000\\1000\\1100\\1110\\1111\end{array}\right]$} & 0.2101\\ \cline{2-2}\cline{4-4}
		& \sr{-0.45cm}{1.1cm} 0.1815 & & 0.2043 \\ \hline\hline
		\multirow{2}{*}[0.25cm]{$\left[\begin{array}{c}0000\\1000\\1100\\1101\\1110\\1111\end{array}\right]$}& \sr{-0.65cm}{1.5cm} 0.2004 & \multirow{2}{*}[0.25cm]{$\left[\begin{array}{c}0000\\1000\\1010\\1100\\1110\\1111\end{array}\right]$} & 0.2021 \\\cline{2-2}\cline{4-4}
		& \sr{-0.65cm}{1.5cm} 0.1962 & & 0.1953 \\\hline
		\end{tabular}\caption{Codes with a large difference between thresholds with and without postselection by Alice on the same code as Bob. The thresholds are indicated to the right of each code and are accurate to within $8.5\times 10^{-4}$. The top threshold corresponds to the states \eqref{eq-filtered_state} without postselection by Alice and the bottom to the states \eqref{eq-filtered_state_Alice} with postselection by Alice on the same code as Bob.}\label{tab-high_difference}
	\end{table}
	
	Note also that all of the codes in Table \ref{tab-high_difference} are nonlinear. In fact, for all the inequivalent linear codes tested (which, apart from the two-code-word repetition codes, comprised only four-code-word codes), the difference between the thresholds was on the order of $10^{-5}$ or less. We also observed the following for the best codes presented in Sec. \ref{subsec-best_codes}. Since the thresholds for the best codes in the $m=2$ and $m=4$ classes are the repetition code thresholds, as mentioned in Sec. \ref{sec-Alice_postselection} the thresholds with and without postselection by Alice are equal. For the six best $m=3$ codes, the threshold difference was on the order of $10^{-5}$, for the two best $m=5$ and $m=6$ codes on the order of $10^{-8}$, and for the single best $m=7$ code on the order of $10^{-11}$.

\end{document}